\documentclass[a4paper,fleqn,usenatbib]{mnras}

\usepackage[T1]{fontenc}
\usepackage{ae,aecompl}
\usepackage{relsize}

\usepackage{graphicx}	
\usepackage{amsmath}	
\usepackage{amssymb}	

\usepackage{longtable}

\usepackage{multirow}

\newcommand\eg{{\it e.g.} }

\newcommand\ie{{\it i.e.} }

\newcommand{\ditto}[1][.4pt]{~\textquotedbl~}

\newcommand{\secondtable}{

\onecolumn

\begin{longtable}{ccc cc cc cc}

\hline

Name&  T$_{\rm eff}$ & Ca/H(e)& $\sigma_{Ca}$& Fe/H(e)& $\sigma_{Fe}$& Ca/Fe& $\sigma_{\rm Ca/Fe}$\\
 & K &  dex & dex & dex & dex & & \\ 

\hline
\\

WD0122-227 $^*$ &8380&      -10.10&        0.10&-8.5&        0.20&        0.03&        0.01 \\ 
WD0446-255 $^*$ &10120&       -7.40&        0.10&-6.9&        0.10&        0.32&        0.10 \\ 
WD0449-259 $^*$ &9850&       -9.10&        0.10&-7.9&        0.20&        0.06&        0.03 \\ 
WD1350-162 $^*$ &11640&       -8.70&        0.10&-7.1&        0.10&        0.03&        0.01 \\ 
WD2105-820 $^*$ &10890&       -8.20&        0.10&-6.0&        0.20&        0.01&        0.00 \\ 
WD2115-560 $^*$ &9600&       -7.40&        0.10&-6.4&        0.10&        0.10&        0.03 \\ 
WD2157-574 $^*$ &7010&       -8.10&        0.10&-7.3&        0.10&        0.16&        0.05 \\ 
WD2216-657 $^*$ &9190&       -9.00&        0.10&-8.0&        0.20&        0.10&        0.05 \\ 
GD 40 $^\alpha$ &15300.0&       -6.90&        0.20&-6.47&        0.12&        0.37&        0.20 \\ 
GD 61 $^\beta$ &17300.0&       -7.90&        0.06&-7.6&        0.07&        0.50&        0.11 \\ 
SDSS J0738+1835 $^\gamma$ &14000.0&       -6.23&        0.15&-4.98&        0.09&        0.06&        0.02 \\ 
PG 1225-079 $^\epsilon$$^\eta$ &10800.0&       -8.06&        0.03&-7.42&        0.07&        0.23&        0.04 \\ 
GD 362 $^\epsilon$ &10500.0&       -6.24&        0.10&-5.65&        0.10&        0.26&        0.08 \\ 
G241-6 $^{\textdaggerdbl}$ &15300.0&       -7.30&        0.20&-6.82&        0.14&        0.33&        0.19 \\ 
HS 2253+8023 $^\eta$ &14400.0&       -6.99&        0.03&-6.17&        0.03&        0.15&        0.01 \\ 
SDSS J124231+522627 $^\theta$ &13000.0&       -6.53&        0.10&-5.9&        0.15&        0.23&        0.10 \\ 
WD 1536+520 $^\iota$ &20800.0&       -5.28&        0.15&-4.5&        0.15&        0.17&        0.08 \\ 
SDSS J0845+2257 $^\kappa$$^\lambda$ &19780.0&       -5.95&        0.10&-4.6&        0.20&        0.04&        0.02 \\ 
PG 1015+161 $^\lambda$ &19200.0&       -6.45&        0.20&-5.5&        0.30&        0.11&        0.09 \\ 
SDSS 1228+1040 $^\lambda$ &20900.0&       -5.94&        0.20&-5.2&        0.30&        0.18&        0.15 \\ 
GALEX 1931+0117 $^\lambda$ &21200.0&       -6.11&        0.04&-4.5&        0.30&        0.02&        0.02 \\ 
G149-28 $^\zeta$ &6020.0&       -8.04&        0.16&-7.41&        0.15&        0.23&        0.12 \\ 
G29-38 $^\rho$ &11800.0&       -6.58&        0.12&-5.9&        0.10&        0.21&        0.08 \\ 
NLTT 43806 $^\zeta$ &5900.0&       -7.90&        0.19&-7.8&        0.17&        0.79&        0.47 \\ 
SDSS J0002+3209 $^\dagger$ &6410.0&       -9.05&        0.18&-7.84&        0.18&        0.06&        0.04 \\ 
SDSS J0006+0520 $^\dagger$ &6790.0&       -9.00&        0.16&-8.39&        0.16&        0.25&        0.13 \\ 
SDSS J0010-0430 $^\dagger$ &6960.0&       -8.38&        0.15&-7.12&        0.15&        0.05&        0.03 \\ 
SDSS J0019+2209 $^\dagger$ &5970.0&       -9.34&        0.16&-8.58&        0.16&        0.17&        0.09 \\ 
SDSS J0044+0418 $^\dagger$ &6050.0&       -9.82&        0.08&-8.71&        0.08&        0.08&        0.02 \\ 
SDSS J0046+2717 $^\dagger$ &7640.0&       -7.65&        0.24&-6.84&        0.24&        0.15&        0.12 \\ 
SDSS J0047+1628 $^\dagger$ &6620.0&       -7.68&        0.18&-6.47&        0.18&        0.06&        0.04 \\ 
SDSS J0108-0537 $^\dagger$ &6010.0&       -8.79&        0.14&-8.08&        0.14&        0.19&        0.09 \\ 
SDSS J0114+3505 $^\dagger$ &6370.0&       -8.51&        0.21&-7.2&        0.21&        0.05&        0.03 \\ 
SDSS J0116+2050 $^\dagger$ &6207.0&       -8.81&        0.08&-7.6&        0.08&        0.06&        0.02 \\ 
SDSS J0117+0021 $^\dagger$ &6800.0&       -8.80&        0.08&-7.6&        0.08&        0.06&        0.02 \\ 
SDSS J0126+2534 $^\dagger$ &5320.0&       -9.95&        0.14&-8.64&        0.14&        0.05&        0.02 \\ 
SDSS J0135+1302 $^\dagger$ &5800.0&       -9.50&        0.11&-8.69&        0.11&        0.15&        0.06 \\ 
SDSS J0143+0113 $^\dagger$ &6700.0&       -8.50&        0.08&-7.3&        0.08&        0.06&        0.02 \\ 
SDSS J0144+1920 $^\dagger$ &6500.0&       -8.50&        0.18&-7.39&        0.18&        0.08&        0.05 \\ 
SDSS J0148-0112 $^\dagger$ &6830.0&       -8.82&        0.20&-7.31&        0.20&        0.03&        0.02 \\ 
SDSS J0150+1354 $^\dagger$ &6310.0&       -7.75&        0.17&-7.24&        0.17&        0.31&        0.17 \\ 
SDSS J0158-0942 $^\dagger$ &5940.0&       -9.52&        0.18&-8.41&        0.18&        0.08&        0.04 \\ 
SDSS J0201+2015 $^\dagger$ &6180.0&       -8.96&        0.14&-8.25&        0.14&        0.19&        0.09 \\ 
SDSS J0252-0401 $^\dagger$ &6950.0&       -8.57&        0.14&-7.46&        0.14&        0.08&        0.04 \\ 
SDSS J0252+0054 $^\dagger$ &7500.0&       -8.35&        0.16&-7.14&        0.16&        0.06&        0.03 \\ 
SDSS J0447+1124 $^\dagger$ &6530.0&       -8.77&        0.19&-8.06&        0.19&        0.19&        0.12 \\ 
SDSS J0512-0505 $^\dagger$ &5563.0&       -8.99&        0.06&-7.79&        0.06&        0.06&        0.01 \\ 
SDSS J0721+3928 $^\dagger$ &6280.0&       -8.90&        0.15&-8.09&        0.15&        0.15&        0.08 \\ 
SDSS J0736+4118 $^\dagger$ &5100.0&       -8.50&        0.12&-7.69&        0.12&        0.15&        0.06 \\ 
SDSS J0741+3146 $^\dagger$ &5592.0&       -9.55&        0.16&-7.5&        0.16&        0.01&        0.00 \\ 
SDSS J0744+1640 $^\dagger$ &4940.0&      -10.24&        0.24&-9.33&        0.24&        0.12&        0.10 \\ 
SDSS J0744+2701 $^\dagger$ &7890.0&       -7.68&        0.17&-6.87&        0.17&        0.15&        0.08 \\ 
SDSS J0744+4408 $^\dagger$ &6370.0&       -8.75&        0.23&-7.54&        0.23&        0.06&        0.05 \\ 
SDSS J0744+4649 $^\dagger$ &5028.0&       -8.36&        0.08&-8.17&        0.08&        0.65&        0.16 \\ 
SDSS J0806+3055 $^\dagger$ &6900.0&       -7.77&        0.23&-7.16&        0.23&        0.25&        0.19 \\ 
SDSS J0806+4058 $^\dagger$ &6808.0&       -8.49&        0.08&-7.49&        0.08&        0.10&        0.03 \\ 
SDSS J0816+2330 $^\dagger$ &7790.0&       -7.48&        0.24&-6.37&        0.24&        0.08&        0.06 \\ 
SDSS J0818+1247 $^\dagger$ &6810.0&       -8.58&        0.25&-7.77&        0.25&        0.15&        0.13 \\ 
SDSS J0823+0546 $^\dagger$ &6019.0&       -9.34&        0.06&-7.36&        0.06&        0.01&        0.00 \\ 
SDSS J0830-0319 $^\dagger$ &6400.0&       -9.10&        0.11&-8.29&        0.11&        0.15&        0.05 \\ 
SDSS J0838+2322 $^\dagger$ &5670.0&       -9.80&        0.09&-9.3&        0.09&        0.32&        0.09 \\ 
SDSS J0842+1406 $^\dagger$ &7160.0&       -8.16&        0.08&-7.3&        0.08&        0.14&        0.04 \\ 
SDSS J0842+1536 $^\dagger$ &6180.0&       -9.47&        0.22&-8.46&        0.22&        0.10&        0.07 \\ 
SDSS J0843+5614 $^\dagger$ &6600.0&       -8.65&        0.16&-7.74&        0.16&        0.12&        0.07 \\ 
SDSS J0851+1543 $^\dagger$ &6300.0&       -8.50&        0.09&-8.2&        0.09&        0.50&        0.14 \\ 
SDSS J0852+3402 $^\dagger$ &5580.0&       -9.00&        0.20&-7.79&        0.20&        0.06&        0.04 \\ 
SDSS J0901+0752 $^\dagger$ &7100.0&       -7.12&        0.11&-6.21&        0.11&        0.12&        0.04 \\ 
SDSS J0902+1004 $^\dagger$ &7250.0&       -8.25&        0.22&-8.19&        0.22&        0.87&        0.63 \\ 
SDSS J0906+1141 $^\dagger$ &6910.0&       -7.90&        0.25&-6.94&        0.25&        0.11&        0.09 \\ 
SDSS J0908+5136 $^\dagger$ &6180.0&       -9.35&        0.10&-8.24&        0.10&        0.08&        0.02 \\ 
SDSS J0913+2627 $^\dagger$ &5210.0&       -9.75&        0.21&-8.64&        0.21&        0.08&        0.05 \\ 
SDSS J0916+2540 $^\dagger$ &5378.0&       -7.48&        0.08&-7.09&        0.08&        0.41&        0.10 \\ 
SDSS J0924+4301 $^\dagger$ &5950.0&       -9.75&        0.26&-8.54&        0.26&        0.06&        0.05 \\ 
SDSS J0925+3130 $^\dagger$ &5810.0&       -9.00&        0.10&-7.99&        0.10&        0.10&        0.03 \\ 
SDSS J0929+4247 $^\dagger$ &6530.0&       -8.46&        0.16&-7.15&        0.16&        0.05&        0.03 \\ 
SDSS J0937+5228 $^\dagger$ &6600.0&       -8.40&        0.09&-7.5&        0.09&        0.13&        0.04 \\ 
SDSS J0939+4136 $^\dagger$ &6310.0&       -8.30&        0.18&-6.79&        0.18&        0.03&        0.02 \\ 
SDSS J0939+5019 $^\dagger$ &5980.0&       -8.25&        0.20&-7.14&        0.20&        0.08&        0.05 \\ 
SDSS J0948+3008 $^\dagger$ &6000.0&       -9.15&        0.14&-8.44&        0.14&        0.19&        0.09 \\ 
SDSS J0956+5912 $^\dagger$ &8800.0&       -7.15&        0.09&-6.14&        0.09&        0.10&        0.03 \\ 
SDSS J1006+1752 $^\dagger$ &5710.0&       -9.45&        0.21&-8.34&        0.21&        0.08&        0.05 \\ 
SDSS J1017+2419 $^\dagger$ &7200.0&       -8.07&        0.13&-6.96&        0.13&        0.08&        0.03 \\ 
SDSS J1017+3447 $^\dagger$ &6150.0&       -9.34&        0.19&-8.33&        0.19&        0.10&        0.06 \\ 
SDSS J1019+2045 $^\dagger$ &5300.0&       -9.36&        0.26&-8.25&        0.26&        0.08&        0.07 \\ 
SDSS J1024+4531 $^\dagger$ &5980.0&       -8.92&        0.17&-8.11&        0.17&        0.15&        0.09 \\ 
SDSS J1033+1809 $^\dagger$ &6070.0&       -8.55&        0.24&-8.04&        0.24&        0.31&        0.25 \\ 
SDSS J1038-0036 $^\dagger$ &7700.0&       -7.85&        0.06&-7.4&        0.06&        0.35&        0.06 \\ 
SDSS J1038+0432 $^\dagger$ &6510.0&       -7.50&        0.16&-6.99&        0.16&        0.31&        0.16 \\ 
SDSS J1040+2407 $^\dagger$ &5750.0&       -8.20&        0.10&-7.59&        0.10&        0.25&        0.08 \\ 
SDSS J1041+3432 $^\dagger$ &7500.0&       -8.20&        0.19&-7.29&        0.19&        0.12&        0.08 \\ 
SDSS J1043+3516 $^\dagger$ &6720.0&       -8.88&        0.12&-7.2&        0.12&        0.02&        0.01 \\ 
SDSS J1055+3725 $^\dagger$ &5600.0&       -8.24&        0.14&-7.83&        0.14&        0.39&        0.18 \\ 
SDSS J1058+3143 $^\dagger$ &6850.0&       -9.02&        0.09&-8.01&        0.09&        0.10&        0.03 \\ 
SDSS J1102+0214 $^\dagger$ &5730.0&       -9.75&        0.10&-8.74&        0.10&        0.10&        0.03 \\ 
SDSS J1103+4144 $^\dagger$ &5850.0&       -9.30&        0.11&-8.04&        0.11&        0.05&        0.02 \\ 
SDSS J1112+0700 $^\dagger$ &7560.0&       -8.53&        0.15&-7.37&        0.15&        0.07&        0.03 \\ 
SDSS J1134+1542 $^\dagger$ &6680.0&       -8.46&        0.24&-7.35&        0.24&        0.08&        0.06 \\ 
SDSS J1144+1218 $^\dagger$ &5434.0&       -9.33&        0.11&-8.37&        0.11&        0.11&        0.04 \\ 
SDSS J1144+3720 $^\dagger$ &7490.0&       -8.17&        0.17&-7.16&        0.17&        0.10&        0.05 \\ 
SDSS J1149+0519 $^\dagger$ &7310.0&       -8.16&        0.12&-7.6&        0.12&        0.28&        0.11 \\ 
SDSS J1150+4928 $^\dagger$ &7210.0&       -8.76&        0.14&-7.65&        0.14&        0.08&        0.04 \\ 
SDSS J1158+0454 $^\dagger$ &5270.0&       -8.69&        0.22&-7.58&        0.22&        0.08&        0.05 \\ 
SDSS J1158+1845 $^\dagger$ &7250.0&       -7.75&        0.15&-6.84&        0.15&        0.12&        0.06 \\ 
SDSS J1158+4712 $^\dagger$ &7650.0&       -8.06&        0.18&-6.85&        0.18&        0.06&        0.04 \\ 
SDSS J1158+5942 $^\dagger$ &6000.0&       -8.98&        0.18&-8.02&        0.18&        0.11&        0.06 \\ 
SDSS J1205+3536 $^\dagger$ &6070.0&       -8.74&        0.16&-7.63&        0.16&        0.08&        0.04 \\ 
SDSS J1211+2326 $^\dagger$ &6450.0&       -8.59&        0.23&-7.28&        0.23&        0.05&        0.04 \\ 
SDSS J1217+1157 $^\dagger$ &6440.0&       -8.93&        0.14&-7.92&        0.14&        0.10&        0.05 \\ 
SDSS J1218+0023 $^\dagger$ &6100.0&       -9.61&        0.09&-8.9&        0.09&        0.19&        0.06 \\ 
SDSS J1220+0929 $^\dagger$ &6640.0&       -8.38&        0.14&-7.37&        0.14&        0.10&        0.05 \\ 
SDSS J1224+2838 $^\dagger$ &5210.0&      -10.00&        0.13&-8.89&        0.13&        0.08&        0.03 \\ 
SDSS J1229+0743 $^\dagger$ &6160.0&       -8.20&        0.13&-7.09&        0.13&        0.08&        0.03 \\ 
SDSS J1230+3143 $^\dagger$ &6510.0&       -9.12&        0.16&-8.21&        0.16&        0.12&        0.06 \\ 
SDSS J1234+5208 $^\dagger$ &7630.0&       -7.40&        0.09&-6.39&        0.09&        0.10&        0.03 \\ 
SDSS J1238+2149 $^\dagger$ &5440.0&       -9.11&        0.22&-8.0&        0.22&        0.08&        0.06 \\ 
SDSS J1245+0822 $^\dagger$ &6360.0&       -8.10&        0.19&-7.59&        0.19&        0.31&        0.19 \\ 
SDSS J1254+3551 $^\dagger$ &6620.0&       -8.96&        0.21&-7.75&        0.21&        0.06&        0.04 \\ 
SDSS J1257-0310 $^\dagger$ &6280.0&       -8.52&        0.23&-7.31&        0.23&        0.06&        0.05 \\ 
SDSS J1259+3112 $^\dagger$ &5840.0&       -9.65&        0.26&-8.14&        0.26&        0.03&        0.03 \\ 
SDSS J1303+4055 $^\dagger$ &6200.0&       -9.02&        0.12&-8.11&        0.12&        0.12&        0.05 \\ 
SDSS J1308+0258 $^\dagger$ &6030.0&       -9.07&        0.22&-7.66&        0.22&        0.04&        0.03 \\ 
SDSS J1316+1918 $^\dagger$ &5350.0&       -9.90&        0.23&-8.99&        0.23&        0.12&        0.09 \\ 
SDSS J1319+3641 $^\dagger$ &7360.0&       -8.60&        0.16&-7.49&        0.16&        0.08&        0.04 \\ 
SDSS J1329+1301 $^\dagger$ &6810.0&       -8.55&        0.14&-7.44&        0.14&        0.08&        0.04 \\ 
SDSS J1330+3029 $^\dagger$ &6100.0&       -8.40&        0.06&-7.3&        0.06&        0.08&        0.02 \\ 
SDSS J1336+3547 $^\dagger$ &6600.0&       -8.50&        0.07&-7.39&        0.07&        0.08&        0.02 \\ 
SDSS J1339+2643 $^\dagger$ &6300.0&       -9.13&        0.07&-8.6&        0.07&        0.30&        0.07 \\ 
SDSS J1340+2702 $^\dagger$ &7855.0&       -6.98&        0.22&-6.27&        0.22&        0.19&        0.14 \\ 
SDSS J1345+1153 $^\dagger$ &6020.0&       -8.10&        0.21&-6.89&        0.21&        0.06&        0.04 \\ 
SDSS J1347+1415 $^\dagger$ &6740.0&       -8.50&        0.11&-7.29&        0.11&        0.06&        0.02 \\ 
SDSS J1350+1058 $^\dagger$ &5120.0&      -10.06&        0.20&-8.75&        0.20&        0.05&        0.03 \\ 
SDSS J1351+2645 $^\dagger$ &5980.0&       -8.04&        0.17&-7.53&        0.17&        0.31&        0.17 \\ 
SDSS J1356+0236 $^\dagger$ &8260.0&       -7.52&        0.15&-6.41&        0.15&        0.08&        0.04 \\ 
SDSS J1356+2416 $^\dagger$ &6030.0&       -9.20&        0.13&-8.54&        0.13&        0.22&        0.09 \\ 
SDSS J1401+3659 $^\dagger$ &6000.0&       -9.80&        0.11&-8.94&        0.11&        0.14&        0.05 \\ 
SDSS J1404+3620 $^\dagger$ &5900.0&       -9.20&        0.08&-8.5&        0.08&        0.20&        0.05 \\ 
SDSS J1405+1549 $^\dagger$ &7150.0&       -8.25&        0.10&-7.14&        0.10&        0.08&        0.03 \\ 
SDSS J1405+2542 $^\dagger$ &5880.0&       -9.50&        0.20&-8.39&        0.20&        0.08&        0.05 \\ 
SDSS J1411+3410 $^\dagger$ &5480.0&       -8.40&        0.24&-7.39&        0.24&        0.10&        0.08 \\ 
SDSS J1421+1843 $^\dagger$ &7300.0&       -7.35&        0.18&-6.19&        0.18&        0.07&        0.04 \\ 
SDSS J1428+4403 $^\dagger$ &6600.0&       -8.98&        0.06&-8.4&        0.06&        0.26&        0.05 \\ 
SDSS J1430-0151 $^\dagger$ &6150.0&       -7.55&        0.16&-6.99&        0.16&        0.28&        0.14 \\ 
SDSS J1445+0913 $^\dagger$ &6580.0&       -7.98&        0.22&-6.77&        0.22&        0.06&        0.04 \\ 
SDSS J1448+1047 $^\dagger$ &6550.0&       -8.85&        0.09&-7.8&        0.09&        0.09&        0.03 \\ 
SDSS J1502+3744 $^\dagger$ &5410.0&      -10.00&        0.12&-8.99&        0.12&        0.10&        0.04 \\ 
SDSS J1518+0506 $^\dagger$ &5020.0&       -9.65&        0.13&-8.74&        0.13&        0.12&        0.05 \\ 
SDSS J1524+4049 $^\dagger$ &5900.0&       -8.90&        0.12&-7.79&        0.12&        0.08&        0.03 \\ 
SDSS J1535+1247 $^\dagger$ &5773.0&       -8.61&        0.05&-7.57&        0.05&        0.09&        0.02 \\ 
SDSS J1542+4650 $^\dagger$ &6060.0&       -8.15&        0.23&-7.04&        0.23&        0.08&        0.06 \\ 
SDSS J1545+5236 $^\dagger$ &5840.0&       -9.19&        0.13&-8.18&        0.13&        0.10&        0.04 \\ 
SDSS J1546+3009 $^\dagger$ &6600.0&       -8.40&        0.12&-7.19&        0.12&        0.06&        0.02 \\ 
SDSS J1549+2633 $^\dagger$ &6290.0&       -9.66&        0.18&-8.25&        0.18&        0.04&        0.02 \\ 
SDSS J1554+1735 $^\dagger$ &6630.0&       -8.60&        0.07&-7.64&        0.07&        0.11&        0.02 \\ 
SDSS J1604+1830 $^\dagger$ &6400.0&       -9.48&        0.12&-8.57&        0.12&        0.12&        0.05 \\ 
SDSS J1616+3303 $^\dagger$ &6400.0&       -8.25&        0.08&-7.14&        0.08&        0.08&        0.02 \\ 
SDSS J1626+3303 $^\dagger$ &6260.0&       -8.87&        0.25&-7.61&        0.25&        0.05&        0.05 \\ 
SDSS J1627+4646 $^\dagger$ &6420.0&       -8.88&        0.20&-7.62&        0.20&        0.05&        0.04 \\ 
SDSS J1636+1619 $^\dagger$ &4410.0&       -9.50&        0.21&-8.79&        0.21&        0.19&        0.13 \\ 
SDSS J1641+1856 $^\dagger$ &5820.0&      -10.30&        0.11&-9.59&        0.11&        0.19&        0.07 \\ 
SDSS J1649+2238 $^\dagger$ &5332.0&       -8.62&        0.18&-7.21&        0.18&        0.04&        0.02 \\ 
SDSS J1706+2541 $^\dagger$ &6640.0&       -9.40&        0.25&-8.89&        0.25&        0.31&        0.25 \\ 
SDSS J2109-0039 $^\dagger$ &6040.0&       -8.78&        0.25&-7.67&        0.25&        0.08&        0.06 \\ 
SDSS J2123+0016 $^\dagger$ &5230.0&      -10.01&        0.17&-8.5&        0.17&        0.03&        0.02 \\ 
SDSS J2157+1206 $^\dagger$ &6100.0&       -9.00&        0.10&-8.09&        0.10&        0.12&        0.04 \\ 
SDSS J2225+2338 $^\dagger$ &7000.0&       -8.81&        0.12&-7.8&        0.12&        0.10&        0.04 \\ 
SDSS J2231+0906 $^\dagger$ &5900.0&       -9.85&        0.09&-8.84&        0.09&        0.10&        0.03 \\ 
SDSS J2235-0056 $^\dagger$ &6340.0&       -8.67&        0.20&-7.26&        0.20&        0.04&        0.03 \\ 
SDSS J2238-0113 $^\dagger$ &6800.0&       -8.89&        0.25&-7.78&        0.25&        0.08&        0.06 \\ 
SDSS J2238+0213 $^\dagger$ &7000.0&       -8.56&        0.23&-7.3&        0.23&        0.05&        0.04 \\ 
SDSS J2304+2415 $^\dagger$ &5060.0&       -9.54&        0.13&-8.93&        0.13&        0.25&        0.10 \\ 
SDSS J2319+3018 $^\dagger$ &7120.0&       -8.53&        0.18&-7.37&        0.18&        0.07&        0.04 \\ 
SDSS J2330+2805 $^\dagger$ &6670.0&       -8.84&        0.20&-7.63&        0.20&        0.06&        0.04 \\ 
SDSS J2340+0124 $^\dagger$ &6200.0&       -8.50&        0.10&-7.9&        0.10&        0.25&        0.08 \\ 
SDSS J2340+0817 $^\dagger$ &5550.0&       -9.00&        0.14&-7.69&        0.14&        0.05&        0.02 \\ 
SDSS J2352+3344 $^\dagger$ &7230.0&       -8.26&        0.20&-7.05&        0.20&        0.06&        0.04 \\ 
SDSS J2357+2348 $^\dagger$ &6030.0&       -9.07&        0.25&-7.76&        0.25&        0.05&        0.04 \\ 
NLTT 1675 $^\phi$ &9999.0&       -9.53&        0.03&-8.63&        0.13&        0.13&        0.04 \\ 
NLTT 6390 $^\phi$ &999.0&      -10.00&        0.04&-8.57&        0.11&        0.04&        0.01 \\ 
NLTT 19686 $^\xi$ &999.0&       -8.70&        0.04&-8.93&        0.14&        1.70&        0.57 \\

\\
\\

\caption{The sample of polluted white dwarfs where both calcium and iron were detected, as used in this work
$^\alpha$\citet{Jura2012}
$^\beta$\citet{Farihi2013} 
$^\gamma$\citet{Dufour2012}
$^\epsilon$\citet{Xu2013}
$^\zeta$\citet{Zuckerman2011}
$^\eta$\citet{Klein2011}
$^\theta$\citet{Raddi2015}
$^\iota$\citet{Farihi2016}
$^\kappa$\citet{Wilson2015} $^\lambda$\citet{Gaensicke2012}
$^\dagger$\citet{Hollands2017}
$^\phi$\citet{Kawka2012}
$^\xi$\citet{Kawka2016}
$^*$\citet{Swan2019}
$^{\textdaggerdbl}$\citet{Jura2012, ZK10}
}

\label{tab:abundances}

\end{longtable}

\twocolumn
}

\title[ ]{ Are exoplanetesimals differentiated? }

\author[A. Bonsor et al.]{
Amy Bonsor$^{1}$\thanks{E-mail: abonsor@ast.cam.ac.uk},  Philip J. Carter$^{2}$, Mark Hollands$^{3}$, Boris T. G{\"a}nsicke$^{3}$, Zo{\"e} Leinhardt$^{4}$ \newauthor and John H. D. Harrison$^{1}$\\ 
$^{1}$Institute of Astronomy, University of Cambridge, Madingley Road, Cambridge, CB3 0HA, UK\\
$^{2}$Department of Earth and Planetary Sciences, University of California Davis, One Shields Avenue, Davis, CA 95616, USA\\
$^{3}$Department of Physics, University of Warwick, Coventry CV4 7AL, UK\\
$^{4}$School of Physics, University of Bristol, HH Wills Physics Laboratory, Tyndall Avenue, Bristol BS8 1TL, UK.
}

\date{Accepted XXX. Received YYY; in original form ZZZ}

\pubyear{2015}

\begin{document}
\label{firstpage}
\pagerange{\pageref{firstpage}--\pageref{lastpage}}
\maketitle

\begin{abstract}

Metals observed in the atmospheres of white dwarfs suggest that many have recently accreted planetary bodies. In some cases, the compositions observed suggest the accretion of material dominantly from the core (or the mantle) of a differentiated planetary body. Collisions between differentiated exoplanetesimals produce such fragments. In this work, we take advantage of the large numbers of white dwarfs where at least one siderophile (core-loving) and one lithophile (rock-loving) species have been detected to assess how commonly exoplanetesimals differentiate. We utilise N-body simulations that track the fate of core and mantle material during the collisional evolution of planetary systems to show that most remnants of differentiated planetesimals retain core fractions similar to their parents, whilst some are extremely core-rich or mantle-rich. Comparison with the white dwarf data for calcium and iron indicates that the data are consistent with a model in which $66^{+4}_{-6}\%$ have accreted the remnants of differentiated planetesimals, whilst $31^{+5}_{-5}\%$ have Ca/Fe abundances altered by the effects of heating (although the former can be as high as $100\%$, if heating is ignored). These conclusions assume pollution by a single body and that collisional evolution retains similar features across diverse planetary systems. These results imply that both collisions and differentiation are key processes in exoplanetary systems. We highlight the need for a larger sample of polluted white dwarfs with precisely determined metal abundances to better understand the process of differentiation in exoplanetary systems.

\end{abstract}


\section{Introduction}

Elements heavier than helium sink below the observable atmospheres of white dwarfs on timescales of days (young, hydrogen-rich or DA white dwarfs) to millions of years (old, helium-rich or DB white dwarfs) \citep{Koester09}. The presence of heavy elements in the atmospheres of $>30\%$ of white dwarfs \citep{Zuckerman03,ZK10, Koester2014} can only be explained by their recent or on-going accretion. The consensus in the literature is that we are observing the accretion of planetary bodies that have survived the star's evolution in an outer planetary system orbiting the white dwarf \citep[\eg][]{JuraWD03, Farihi_review, Veras_review}. Dynamical instabilities following stellar mass loss can scatter planetary bodies onto star-grazing orbits \citep{DebesSigurdsson, bonsor11, Veras_twoplanet_2013, debesasteroidbelt} where they are disrupted by strong tidal forces \citep{Veras_tidaldisruption1}. Fragments of disrupted bodies are accreted onto the star, with observations of gas and dust tracing this accretion in action (see \cite{Farihi_review} for a recent review).

The abundances observed in the atmospheres of white dwarfs provide unique insights regarding the composition of exoplanetary building blocks; the planetesimals accreted by the white dwarfs. Whilst most white dwarf pollutants exhibit abundances that are broadly similar to rocky planets \citep{juraYoung2014}, a handful show the presence of volatiles, including oxygen and nitrogen \citep[\eg][]{Raddi2015,Farihi2011_water, Xu2017}. High abundances of refractory species, such as calcium and titanium, have led to the suggestion that some white dwarf pollutants experienced high temperature processing, similar to meteorites from the inner solar system, for example, G29-38 \citep{Xu2014a}. Extreme abundances of either siderophile (core-loving) species \ie iron or lithophile (rock-loving) species, such as calcium, magnesium or silicon, have led to the suggestion that these polluted white dwarfs have accreted a fragment of a larger body that differentiated into a core and mantle. For example, for SDSS J0845+2257 \citep{Wilson2015}, the high iron abundance could be explained by the accretion of a planetesimal stripped of its mantle. In order to explain the observed abundances, not only must the planetary bodies differentiate, but collisions must be sufficiently catastrophic that at least some fragments have extreme compositions \eg core-rich.


Both collisions and differentiation are common features of our asteroid belt. Samples of differentiated bodies arrive to Earth as meteorites, most famously the iron meteorites. The range of different spectral classifications for asteroids could be explained in part by their differentiation and collisional evolution \citep{Burbine2002}. The budget of short-lived radioactive nuclides in the Solar System, including $^{26}$Al, is sufficient to differentiate bodies larger than $>10$\,km \citep{Urey1955, GhoshMcSween1998}, and there is a growing suite of evidence that differentiation occurred early \citep{Kleine2005, Schersten2006, Kruijer2014}. We note here that there is sufficient potential energy imparted during formation alone to differentiate bodies larger than around 1,000\,km, without the need for short-lived radioactive nuclides \citep[\eg][]{Davison2010,Elkins-Tanton2011}. In fact, the differentiation of planetary building blocks influences the composition of the terrestrial planets, most notably the budgets of highly siderophile elements \citep{Rubie2011, Rubie2015, Fischer2017}, but also potentially the bulk composition \citep{bonsorleinhardt, Carter2015}. However, the budget of short-lived radioactive nuclides in exoplanetary systems is greatly debated \citep[\eg][]{Young2014, Lichtenberg2016, Gounelle2015, Boss2010, Gritschneder2012} and it is not clear how widespread the differentiation of small exoplanetary bodies is. In fact, \cite{Jura2013} previously used the white dwarf observations to suggest that the Solar System's abundance of $^{26}Al$ was not so unusual.

Collisions between differentiated bodies can lead to fragments with a diverse range of compositions. Simulations of disruptive collisions produce fragments with a range of compositions, including those dominated by core material or mantle material \citep{Marcus2010, bonsorleinhardt, Carter2015}. Mercury could be a collision fragment dominated by core material; stripped of its mantle in a high velocity collision \citep{Benz1988}. The merging of planetary cores and the stripping of mantles are common features of high velocity collisions \citep{Benz1988,Marcus2009, Landeau2016}. Similar processes have been hypothesised to occur in exoplanetary systems \citep{Marcus2009}. In this work we consider how the differentiation and collisional evolution of planetesimal belts may influence the compositions of planetary bodies accreted by white dwarfs.

With the growing number of {\it polluted }white dwarfs where both lithophile (\eg calcium) and siderophile (\eg iron) elements have been detected, it is now possible to assess the population of exoplanetary systems as a whole. We hypothesise that all polluted white dwarfs have accreted planetesimals from outer planetesimal belts that have survived the star's evolution. These outer planetesimal belts are commonly observed around main-sequence stars and known to be collisionally active due to the large quantities of small dust continuously replenished in collisions between larger bodies \citep{wyattreview, Hughes2018}. Collisions between planetesimals that have differentiated to form a core and mantle should lead to a spread in total abundances of siderophile (\eg iron) and lithophile (\eg calcium) species. If polluted white dwarfs sample this distribution, it will be reflected by the spread in their observed calcium and iron abundances.

The aim of this work is to collate as large a sample as possible of polluted white dwarfs where both calcium and iron are detected and use these, compared to models for the collisional evolution of differentiated planetesimals, to infer the prevalence of differentiation in exoplanetesimals. Do most planetary systems have planetesimals that differentiate, or is the differentiation of planetesimals a unique feature of the Solar System?

We start by summarising the aims and approach of the paper in \S\ref{sec:aims}, followed by the observational data sample in \S\ref{sec:data}. In \S\ref{sec:stars}, we firstly consider the possibility that the distribution of abundances could be explained by a range of initial abundances for the planet forming material.
 In \S\ref{sec:diff} we compare the white dwarf observations to the results of the simulations and investigate the frequency of differentiation in the observed white dwarf planetary systems. In \S\ref{sec:over} we discuss the overabundance of polluted white dwarfs with above average Ca/Fe ratios and suggest that this trend may be related to the temperatures experienced by the planetesimals. In \S\ref{sec:discussion} we discuss what can be concluded by the currently available data and models, before summarising our conclusions in \S\ref{sec:conclusions}.

\section{Aims and Approach}
\label{sec:aims}
In this work we aim to investigate what the observed population of polluted white dwarfs can tell us about how frequently exoplanetesimals are differentiated. We do this by comparing the observed population to a model population, developed from the results of simulations. The N-body simulations follow the collisional evolution of planetary systems, tracing the fate of core-like and mantle-like material. We use collision simulations to predict the population of fragments that accrete onto white dwarfs and compare to the observed population of white dwarf pollutants. 

Calcium and iron are both commonly observed in polluted white dwarfs, whilst being a pair of lithophile and siderophile elements that behave differently during differentiation. In addition to which both elements sink at relatively similar timescales through the white dwarf atmosphere \citep{Koester09}, which means that the observed abundances should match those of the material accreted onto the star and do not need to be adjusted to take into account differential sinking. The ratio of Ca to Fe should remain unchanged even if the sinking timescales change, for example in hot DA white dwarfs due the onset of thermohaline (fingering) convection sinking timescale may be significantly longer \citep{Deal2013,Zemskova2014,Wachlin2017, Bauer2018, Bauer2019} or accreted material is mixed much deeper into the white dwarf \citep{Cunningham2019}.

Ca and Fe are, therefore, a very useful pair of species for diagnosing the levels of differentiation in exoplanetesimals. We collate as large an observational sample as possible of white dwarfs where both calcium and iron have been detected and compare these to the model predictions. We utilise the cumulative distribution of Ca/Fe ratios, ${\bf k}$, to compare the model to the observations. For the observations, this essentially equates to a list of observed Ca/Fe values in ascending order, each with an associated error, $\sigma_{\rm Ca/Fe}$. We, therefore, define $X_{\rm {Ca/Fe}}^{\rm obs}({\bf k})$ as the value of Ca/Fe for which a fraction ${\bf k}$ of the observed sample have a lower Ca/Fe measurement, or in other words a fraction ${\bf k}$ of the sample have an observed Ca/Fe ratio lower than $X_{\rm {Ca/Fe}}^{\rm obs}({\bf k})$. In a similar manner, a fraction ${\bf k}$ of the model population are predicted to have a lower Ca/Fe value than $X_{{\rm Ca/Fe}}^{\rm model}({\bf k})$. We can, thus, assess the quality of the model fit using a reduced chi-squared of: 
\begin{equation}
\chi_{{\rm model}}^2 = \frac{1}{{\rm N}_{\rm WD}}{\mathlarger{‎‎\sum}}_{k=0}^{1} \left( \frac{X_{{\rm Ca/Fe}}^{\rm model}({\bf k})- X_{{\rm Ca/Fe}}^{\rm obs}({\bf k})}{2 \sigma_{\rm Ca/Fe}({\bf k})}\right)^2,
\label{eq:chisquared}
\end{equation}
where ${\rm N}_{\rm WD}$ is the number of white dwarfs in the sample.

In order to determine the most likely values of the model parameters, we use a Bayesian framework. The posterior probability distribution ($p(\theta|M_i, D)$) of the model parameters, $\theta$, given the model, $M_i$ and the data, $D$, is proportional to the likelihood of the data, given the model and parameters, $\mathcal{L}(D|\theta, M_i)$ and the prior on the model parameters $p(\theta|M_i)$ (see \S~\ref{sec:fdiff} and \S~\ref{sec:hight}). In order to answer the question of whether exoplanetesimals are differentiated, the key model parameter ($\theta$) is the fraction of exoplanetesimals that are differentiated, $f_{\rm diff}$. The Markov Chain Monte Carlo (MCMC) fitting routine \citep{emcee} is used to maxmimise this likelihood function in order to find posterior distribution for each model parameter,  assuming a likelihood of the form :
\begin{eqnarray}
\label{eq:like}
 &\mathcal{L}(X_{\rm Ca/Fe}^{\rm obs}|\theta, M_i)=\\ \nonumber
&\prod\limits_{k=0}^{1} \frac{1}{\sqrt{2\pi}\sigma_{\rm Ca/Fe}({\bf k})}  \, {\rm exp} \left(- \left( \frac{X_{\rm Ca/Fe}^{\rm model}({\bf k})- X_{\rm Ca/Fe}^{\rm obs}({\bf k})}{2 \sigma_{\rm Ca/Fe}({\bf k})}\right)^2\right)\\\nonumber
&ln  \, \mathcal{L}( X_{\rm Ca/Fe}^{\rm obs}|\theta, M_i)  = \\\nonumber
&  - {\mathlarger{‎‎\sum}}_{k=0}^{1}\left(\ln( \sqrt{2 \pi} \,\sigma_{\rm Ca/Fe}({\bf k}))  +\left(\frac{ X_{\rm Ca/Fe}^{\rm obs}({\bf k}) - X_{\rm Ca/Fe}^{\rm model}({\bf k})}{2 \sigma_{\rm Ca/Fe}({\bf k})}\right)^2 \right).
\end{eqnarray}


\section{ Observational Data }

\label{sec:data}

\begin{figure}
\includegraphics[width=0.48\textwidth]{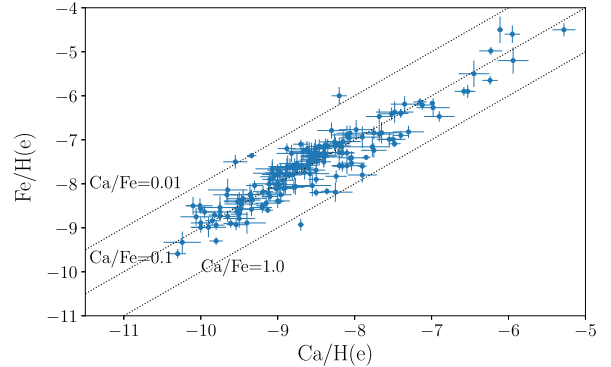}
\caption{The calcium and iron abundances measured in the 179 white dwarfs plotted in the sample. Lines of constant Ca/Fe ratio are over-plotted as dotted lines.   }
\label{fig:obs}
\end{figure}

The observational sample are collated from the literature, some are the most highly polluted white dwarfs where multiple species have been detected \citep{Jura2012,Farihi2013,Dufour2012,Gaensicke2012,Klein2011, Xu2013,Zuckerman2011,Klein2011,Raddi2015,Farihi2016,Wilson2015,Hollands2017,Kawka2012,Kawka2016,Swan2019}, whilst most are cool ($T_*<9,000$K) DZs from \cite{Hollands2017, Hollands2018}. Errors on the measured abundances are taken from the literature, where available ($\sigma_{\rm Ca}$ is the error on $10^{\rm Ca/H(e)}$ and $\sigma_{\rm Fe}$ on $10^{\rm Fe/H(e)}$). We note that quoted errors are often conservative, with an attempt to fold in uncertainities on the atomic data, as well as uncertainities derived from models of the white dwarf atmospheres. Standard error propagation is used to find the error on the Ca/Fe ratio, assuming that the errors on the Ca and Fe abundances are independent: $\sigma_{\rm Ca/Fe} =\frac{10^{\rm Ca/H(e)}}{10^{\rm Fe/H(e)}} \ln(10) ( \sigma_{\rm Ca}^2 + \sigma_{\rm Fe}^2)^{1/2}$. This may not be valid as abundances in Ca and Fe may be correlated. For the \cite{Hollands2017} sample, errors are estimated assuming that they are a sum of systematic errors plus the statistical uncertainty on the abundances, where the systematic errors for Ca and Fe are taken to be $\sigma_{\rm Ca, sys}=\sigma_{\rm Fe, sys}=0.05$ dex and $\sigma_{\rm Ca}^2 = \sigma_{\rm Ca, sys}^2 + \left(\frac{1.27}{(S/N)}\right)^2$, and an equivalent equation for $\sigma_{\rm Fe}^2$, where $S/N$ is the mean spectral signal-to-noise ratio between 4500--5500\AA, taken from (Hollands private communication) and the scaling factor $1.27$ is a conservative estimate, based on the weakest lines in the noisiest spectra matching the maximum error on an individual element detection \citep{Hollands2017}. For those white dwarfs with low signal to noise observations, particularly from \cite{Hollands2017}, the Ca and Fe abundances are poorly known and do not provide information regarding differentiation, we, therefore, focus on a sample of white dwarfs where $S/N>5$ in this work (as suggested by \cite{Hollands2018}), which includes 179 white dwarfs. 
Fig.~\ref{fig:obs} plots the observed Ca and Fe abundances, alongside associated errors. Table~\ref{tab:abundances} lists the Ca and Fe abundances of the sample with associated errors, stellar temperatures and references for all measurements.

\section{A range of initial abundances in the planet-forming material }
\label{sec:stars}

\begin{figure}
\includegraphics[width=0.48\textwidth]{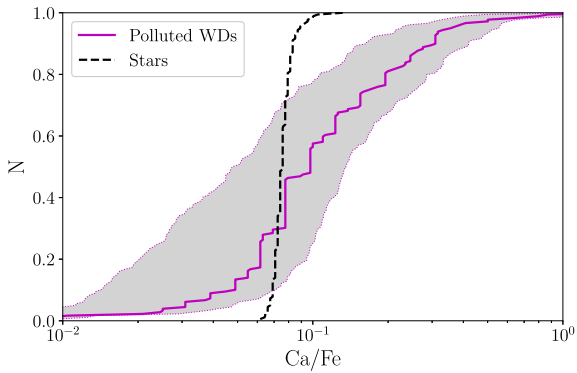}
\caption{The cumulative distribution of Ca/Fe ratios observed in the white dwarf pollutants (magenta solid line, plus grey shaded region indicating $1-\sigma$ errors), compared to the cumulative distribution of Ca/Fe abundances predicted for the scenario in which no exoplanetesimals differentiate (stars: black dashed line). }
\label{fig:cumdist}
\end{figure}

In the scenario that no exoplanetesimals differentiate, a narrow range in the Ca/Fe ratios of exoplanetesimals is expected resulting from the narrow range of Ca/Fe ratios found in the initial conditions of the material from which these planetesimals formed. If we consider that stars and planetary systems form out of the same material, then we can assume that the range of compositions of nearby stars will be fairly similar to the range of initial compositions present in their planetary systems. We can easily determine the compositions of nearby stars, whereas determining the initial compositions of their planetary systems is challenging. We, therefore, consider the range of Ca/Fe ratios found in a sample of nearby stars to be a good proxy for the potential spread expected for undifferentiated, pristine exoplanetesimals. We compare the cumulative distribution of Ca/Fe ratios found in a sample of nearby FGK stars taken from \cite{Brewer2016} to the population of polluted white dwarfs. FGK stars are used as they are more likely to have formed at similar times, thus, with potentially similar compositions, to the progenitors of the white dwarfs considered. Fig.~\ref{fig:cumdist} shows the cumulative distribution of Ca/Fe ratios observed in polluted white dwarfs (magenta line), along with a corresponding error range shown in grey. This is calculated by considering the cumulative distribution that would occur if all white dwarfs in the sample were measured to have their measured Ca/Fe (de)increased by one sigma. This is compared to the distribution of Ca/Fe ratios seen in nearby stars (black line). The polluted white dwarfs show a much broader range of Ca/Fe ratios. We find a reduced $\chi_{\rm stellar}^2=3.5$, indicating a relatively poor fit. If we consider those 35 white dwarfs with $S/N>20$, $\chi_{\rm stellar}^2=12.5$ and this is clearly a bad model for the observational data.



\section{Collisions between differentiated planetesimals }
\label{sec:diff}

Planetesimal belts evolve over many billions of years to the white dwarf phase. Debris discs around main-sequence stars provide evidence that planetesimal belts are collisionally active \citep{wyattreview}. If the planetesimals have differentiated to form a core and a mantle, collisions between these differentiated planetesimals lead to fragments with a range of compositions and, of particular relevance here, a range of Ca/Fe ratios \citep{Marcus2009,Marcus2010, bonsorleinhardt, Carter2015}. The scattering and accretion of planetesimals from planetary systems that have survived the star's main-sequence evolution to the white dwarf phase is thought to be a dynamical process \citep{Veras_review} and thus, broadly independent of a body's collisional or geological history. Thus, we anticipate that if the exoplanetesimals accreted by white dwarfs originate from a system where both collisions and differentiation have occurred, they should randomly sample this distribution of Ca/Fe ratios.

The aim here is to produce a model population which predicts the distribution of Ca/Fe ratios following collisional evolution and random sampling of the collision fragments by the white dwarfs. The exact distribution of Ca/Fe ratios will depend upon the precise architecture, in particular orbital location of material and evolution timescales, of individual systems.  Given that it is not computationally feasible to consider the full range of system architectures, we use a single system as a proxy for all systems. We base our results on simulations of our own Solar System, early in its evolution, namely considering a planetesimal belt in the region around Earth's orbit. Whilst this system goes on to form some proto-planets, it leaves behind a population of collision remnants, produced from a range of destructive and constructive collisions. We use these remnants are a proxy for the bodies accreting onto the white dwarfs.

We hypothesise that if the distribution of Ca/Fe ratios in collision fragments is broadly similar across all planetary systems, the result of the polluted white dwarfs sampling a range of systems will be similar to our model in which a single system is sampled. Crucially, we note that even if the distribution of Ca/Fe ratios exhibits stark differences between individual systems, these will be very difficult to disentangle from the general population and therefore, as a first approach, we can derive significant insights from our model. 

In particular, we anticipate that similar evolution will occur on longer timescales further from the star. We simulate a massive belt (see below), close to the star ($\sim 1 $au) such that the collisional evolution occurs rapidly. However, if we consider that collisions are dominated by catastrophic collisions, \cite{Wyatt07, wyattreview} estimate that the collisional timescale is proportional to the belt orbital radius, $r$ as $r^{13/3}$, such that the evolution occurring in a belt at 1au in 10Myr, occurs at 5au in approximately 10Gyr. Thus, we anticipate that the distribution from a close-in system on short timescales can be used as a proxy for the distribution from a system further out on longer timescales, potentially more relevant to planetary systems around white dwarfs. We note here that most collisional evolution will occur whilst the planetary systems are on the main-sequence, whilst some evolution may continue during the white dwarf phase, at which point the system will have expanded in orbital radii, potentially by a factor of $\sim 3$ due to the reduced stellar mass. The use of a single system at a single epoch to represent the sampling of a range of systems on a range of timescales by the white dwarfs can be justified if the initial evolution of the Ca/Fe distribution is more dramatic than the evolution on longer collision timescales. We will discuss the time dependence of the distribution of Ca/Fe ratios in remnant planetesimals in further detail in \S\ref{sec:limitations}.

To summarise, whilst we hypothesise that the white dwarfs are polluted by planetesimals resulting from a wide range of different planetary systems with different architectures and collision histories, we use a single system as a proxy to predict the distribution of Ca/Fe ratios in the population of planetesimals polluting white dwarfs. This assumes that the generic form of the predicted distribution of Ca/Fe ratios can be applied across a wide range of systems and we, therefore, try not to pin our conclusions down to the specific details of the distribution predicted here.

\subsection{Simulations}
\label{sec:sims}
The N-body simulations were performed using a state-of-the-art N-body code, {\it pkdgrav}. The collisional evolution of a planetesimal belt is followed, taking into account both destructive and accretional collisions. The fate of collision fragments are tracked using the EDACM collision model \citep{leinhardt2015}. Every planetesimal is assumed to start with a given size and initial core mass fraction. The fate of core and mantle material during collisions is tracked separately using prescriptions based on simulations of \citep{Marcus2009,Marcus2010}. Full details of the code and simulations can be found in \cite{leinhardt2015, bonsorleinhardt, Carter2015}. The simulations were originally designed to focus on terrestrial planet formation in the inner regions of planetary systems and, therefore, lead to the formation of several proto-planets. For the purposes of this work we focus on the collisionally evolved population of fragments that remain at the end of the simulations, only considering bodies with a mass less than $0.1M_\oplus$. This population will have undergone similar collisional evolution to that which might occur in a planetesimal belt. The formation of proto-planets can {\it stir} the smaller fragments, inciting further collisional evolution, in a similar manner as is thought to occur in outer debris discs either due to self-stirring or planet stirring \citep{wyattreview, alex, grantstirring}. For a planetesimal belt significantly further from the star, or significantly less dense, the evolution seen in these simulations is representative of the evolution that would occur on significantly longer timescales, timescales that would be computationally infeasible to simulate. Each simulation takes about one month to run. Two types of collision simulations are considered, the first considers Earth formation in a {\it Calm} scenario and was repeated 5 times, whilst the second considers a more specialised scenario involving Earth's formation, but this time including Jupiter's Grand Tack. Full details of both scenarios can be found in \cite{Carter2015}:
\begin{itemize}
\item{ {\bf  Calm:} The evolution of 100,000 planetesimals with an initial radius of between 196\,km and 1530\,km in a belt of $2.5\,M_\oplus$ between 0.5 and 1.5\,au is followed for 20\,Myr\footnote[1]{This is the effective time elapsed, which differs from the time for which the simulation was run by a correction to take into account the expanded radii particles used in order to speed the computation by a factor of 6. See \cite{Carter2015} for more details}. All planetesimals start with an initial core mass fraction of $C_f(0)=0.35$. These are equivalent to Simulations 8--11 ({\it Calm} 1-4), as labelled in \cite{Carter2015}). The remaining difference between the simulations depends on the exact laws used to determine the fate of core and mantle material following collisions. }


\item{ {\bf Grand Tack ({\it GT}):} 
 The evolution of 10,000 planetesimals with an initial radius of between 528\,km and 2250\,km in a belt of $4.85M_\oplus$ between 0.5 and 3.0 \,au is followed for 20\,Myr \footnotemark[1]. The migration of a Jupiter mass planet in a so-called Grand Tack is included, where Jupiter starts at 3.5\,au, migrates inwards to 1.5\,au and then back outwards to finish at 5.2\,au. Gas drag is included with a surface density profile based on hydrodynamical simulations of giant planets embedded in a disc \citep{MorbidelliCrida}. The effect of Jupiter's grand tack is essentially to increase collision velocities, in a similar manner as might occur later in the evolution of a planetary system due to other effects, including stirring by a giant planet. (Simulation 28 in \cite{Carter2015}.}

\end{itemize}
In both simulations, a range of collision fragments are created following each (partially) destructive collision. Fragments that fall below the minimum resolution limit (196\,km for the {\it Calm} simulations) are added to the mass in unresolved dust. This debris is distributed in a series of circular annuli each 0.1\,au wide, with the debris placed in the annulus corresponding to the location of the collision that produced it. This material is assumed to have circular Keplerian orbits. The unresolved debris is accreted onto all planetesimals at each time-step, and the fraction of core and mantle material in the unresolved debris is tracked. Given that mantle stripping is common for small fragments, this unresolved debris tends to be dominated by mantle material and leads to all fragments accreting extra mantle-like material.

\subsubsection{Distribution of core mass fractions}

\begin{figure}

\includegraphics[width=0.48\textwidth]{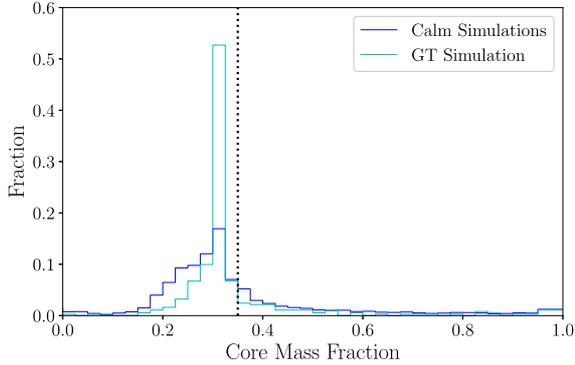}
\caption{The distribution of core mass fractions for fragments with masses less than $0.1M_\oplus$ left at the end of the N-body simulations (see\S\ref{sec:sims}). The 4 {\it Calm} simulations are averaged for clarity. The dotted line indicates the initial core mass fraction of $C_f(0)=0.35$ given to planetesimals. }
\label{fig:hist_sims}
\end{figure}

In this work we are interested in the collisionally evolved fragments and their range of potential abundances. We, therefore, focus on the distribution of core mass fractions of planetesimals, with masses less than $0.1M_\oplus$, that survive to the end of the simulations. Fig.~\ref{fig:hist_sims} shows these distributions. All the simulations display two key features that were also seen in other similar simulations with different initial parameters. Firstly, the distribution of core mass fractions is peaked close to the initial core mass fraction of the planetesimals. Secondly, the simulations produce a wide range of fragments, spread between core-rich and mantle-rich fragments, with the potential for some fragments that are extremely core or mantle-rich.  

In these simulations the peak in the distribution is always shifted slightly towards mantle-rich fragments. This can be attributed to two effects. Firstly, the prescription used in \cite{Carter2015} favours the accretion of cores by the largest remnants following a collision. This leads to the core material being predominantly found in the larger, less numerous remnants, such that most remnants tend towards being more mantle-rich. Secondly, most bodies have grown by accreting the predominantly mantle-rich, unresolved fragments, which fill the orbital parameter space around the original bodies. Whether this is a realistic effect, or a result of the limited resolving power of the simulations that can only track planetesimals down to around 200\,km in size (see \S\ref{sec:sims} for details) remains unclear. The main difference between the Calm simulations and the Grand Tack simulations is that in the Grand Tack simulations the core-mass distribution is more highly peaked. Jupiter excites the planetesimals such that there are more disruptive collisions, which produce more debris. As collisions are more disruptive this debris contains more core material and thus, the accretion of this debris leads to a population of fragments with core mass fractions very close to the average values.

In order to compare the simulation results to the white dwarf observations, we convert the distribution of core mass fractions to a distribution of Ca/Fe ratios in the simplest manner possible, by assuming that the mass fraction of Ca and Fe in the planetesimal's core and mantle are the same as in bulk Earth, respectively. This means that: 


\begin{equation}
\left(\frac{\mathrm Ca}{\mathrm Fe}\right)_{\rm sims} =  \left(\frac{A_{\rm Fe}}{A_{\rm Ca}}\right) \left( \frac{(1-C_f){\rm Ca}_{\rm mantle}^\oplus  }{ C_f {\rm Fe}_{\rm core}^\oplus + (1-C_f) {\rm Fe}_{\rm mantle}^\oplus }\right),
\label{eq:cf}
\end{equation}
where $C_f$ is the core mass fraction of the planetesimal, $A_{\rm Fe}$ and $A_{\rm Ca}$ are the atomic weights for iron and calcium, ${\rm Ca}_{\rm mantle}^\oplus=26.1\times 10^{-3}$ the mass fraction of Ca in Earth's mantle, ${\rm Fe}_{\rm core}^\oplus=0.85$ the mass fraction of iron in the core, and ${\rm Fe}_{\rm mantle}^\oplus=0.063$ (all values from \cite{Palme2003treatise}). The simulations predict the existence of some almost pure core fragments. These will have so little calcium that they are unlikely to be detected and we, therefore, remove from the model population all fragments with core masses higher than $C_f>0.826$ or Ca/Fe$<8.9 \times 10^{-3}$, where Ca/Fe$=8.9 \times 10^{-3}$ is the lowest observed Ca/Fe ratio in any polluted white dwarf in the sample.

We consider that the distribution of core mass fractions produced by the collisional evolution seen in these simulations will be fairly typical of collisional evolution in most planetesimal belts, although the exact details will clearly vary depending on the exact collisional and dynamical history of the individual planetary system.


\begin{figure}
\includegraphics[width=0.48\textwidth]{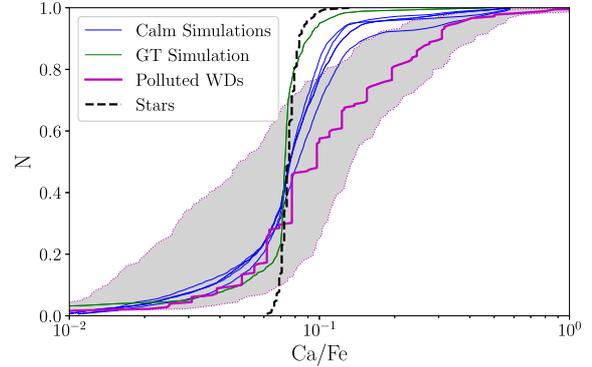}
\caption{The same as Fig.~\ref{fig:cumdist} now including the cumulative distribution of Ca/Fe abundances predicted from the simulations, the blue solid lines show the {\it Calm} simulations and the green solid line the {\it GT} simulations. }
\label{fig:cumdist_sims}
\end{figure}

\begin{figure}
\includegraphics[width=0.48\textwidth]{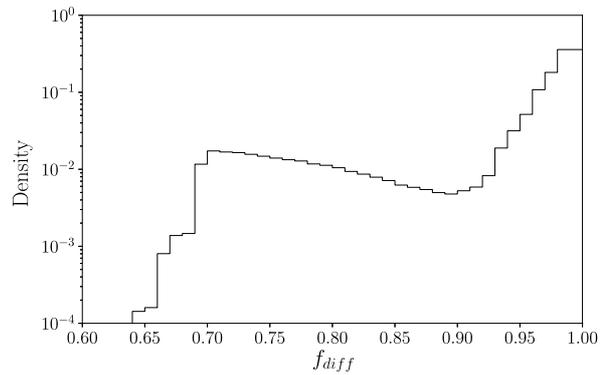}
\caption{The posterior distribution for the fraction of exoplanetesimals that are differentiated, $f_{\rm diff}$, considering only observations with low Ca/Fe (Ca/Fe$<0.07$). Whilst the most likely solution is that almost all exoplanetesimals are differentiated, a good fit to the data can be found with as low as 65\% of exoplanetesimals being differentiated. }
\label{fig:fdiff_coreonly}
\end{figure}


\subsection{The fraction of exoplanetesimals that are differentiated}
\label{sec:fdiff}
 Not all bodies that accrete onto white dwarfs will be fragments of differentiated planetesimals. In some planetary systems, small exoplanetesimals may not differentiate and these may be the planetesimals that accrete onto white dwarfs. In our Solar System, some bodies that formed early in the inner system show clear evidence for differentiation (\ie iron meteorites), whilst other asteroids remain unaltered, such as the chondritic meteorites. In fact, the exact frequency of differentiated bodies within our asteroid belt remains a subject of debate \citep{DeMeo2015}. In exoplanetary systems, there may not have been a sufficient heat budget to lead to differentiation (\eg short-lived radioactive nuclides), or planetesimals may have formed too late to take advantage of this. We, therefore, introduce a parameter $f_{\rm diff}$, which represents the fraction of exoplanetesimals accreted by white dwarfs that are fragments of differentiated bodies. We consider this to be a good proxy for the fraction of exoplanetary systems in which planetesimals are differentiate. The model is now defined as: 
\begin{equation}
k^{\rm diff}({\bf X_{\rm Ca/Fe}}) = f_{\rm diff}\, k^{\rm sims}({\bf X_{\rm Ca/Fe}}) + (1-f_{\rm diff})\,  k^{\rm stars}({\bf X_{\rm Ca/Fe}}),
\end{equation}
where $k^{\rm diff}({\bf X_{\rm Ca/Fe}})$ is the cumulative distribution of Ca/Fe ratios predicted by the model, which is found as a sum of the cumulative distribution of Ca/Fe ratios predicted from the simulations, $k^{\rm sims}({\bf X_{\rm Ca/Fe}})$ and the cumulative distribution of Ca/Fe ratios predicted for the initial conditions at the start of planet formation, $k^{\rm stars}({\bf X_{\rm Ca/Fe}})$,  which are both a function of the Ca/Fe ratio, ${\bf X_{\rm Ca/Fe}}$. A 1-D interpolation is used to convert between $k^{\rm diff}({\bf X_{\rm Ca/Fe}})$ and $ X_{\rm Ca/Fe}^{\rm diff}({\bf k})$.  The MCMC fitting routine \citep{emcee} is used to maximise the likelihood function (Eq.~\ref{eq:like}) in order to find the posterior distribution of $f_{\rm diff}$, assuming a uniform prior in which $0<f_{\rm diff} <1$. The best fit value is very close to 1, with $f_{\rm diff} = 99^{+0.8}_{-1.8}$\% averaged over all {\it Calm} simulations. In other words, the data is consistent with almost all white dwarf pollutants being the fragments of differentiated planetesimals.



Fig.~\ref{fig:cumdist_sims} shows the cumulative distribution of Ca/Fe ratios predicted by the model with differentiation, compared to the observed population. The uncertainties in the observed abundances are such that this model is consistent with the observations ($\chi^2<2$ Eq.~\ref{eq:chisquared}). In fact, such a small $\chi^2$ may be indicative that the errors on the abundances are in some cases overestimated. In most cases, quoted errors are very conservative, taking into account both potential errors on the atomic data and white dwarf atmosphere models.

By comparing the likelihoods (Table~\ref{tab:model}), we can see that a model in which all planetesimals are differentiated is significantly more likely as an explanation for the data than a model in which the range of Ca/Fe ratios is explained only by the small variation in the initial conditions for planet formation (no exoplanetesimals are differentiated).

However, whilst the model is a reasonable fit to the observations at low Ca/Fe ratios, it is less successful at high Ca/Fe ratios. The reasons for this will be discussed in detail in the next section, however, we consider those observations with low Ca/Fe ratios (Ca/Fe $<0.07$) as a useful sample for better constraining the fraction of planetesimals that are differentiated, $f_{\rm diff}$. Ca/Fe$=0.07$ is the median of the distribution of Ca/Fe in nearby stars, whilst also being equivalent to a core mass fraction of $C_f=0.35$, assuming typical parameters for bulk Earth (Eq.~\ref{eq:cf}). Maximising the likelihood (Eq.~\ref{eq:like}) considering only low Ca/Fe ratios (Ca/Fe$<0.07$) finds a posterior distribution of $f_{\rm diff}$ as shown in Fig.~\ref{fig:fdiff_coreonly}. Whilst all exoplanetesimals being differentiated ($f_{\rm diff}=1.0$) remains the most likely model, a good fit to the data is found even when the fraction of exoplanetesimals that are differentiated is as low as 65\%.








\begin{figure}

\includegraphics[width=0.48\textwidth]{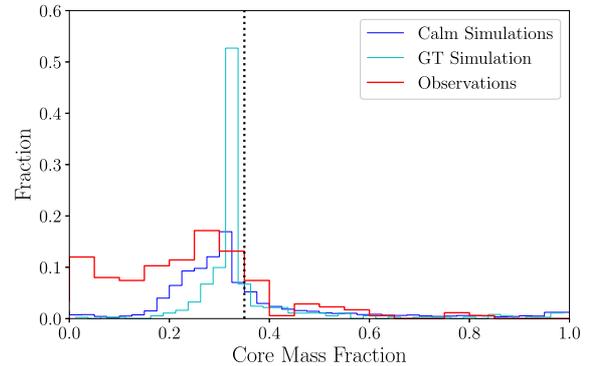}
\caption{The overabundance of white dwarfs with high Ca/Fe ratios, as discussed in \S\ref{sec:over}, is apparent when only the Ca/Fe ratio is used to predict the core mass fractions of the bodies accreted by the white dwarf sample (Eq.~\ref{eq:cf}), shown in the figure compared to the distribution of core mass fractions from the simulations (see \,\S\ref{sec:sims}). In \S, we discuss alternative explanations to explain the overabundance of apparently core-rich fragments, derived from the overabundance of polluted white dwarfs with high Ca/Fe. }
\label{fig:hist_cf}
\end{figure}

\section{The overabundance of polluted white dwarfs with high C\lowercase{a}/F\lowercase{e} ratios}
\label{sec:over}
There are more polluted white dwarfs with high Ca/Fe ratios in the observed population than in the model population in which all exoplanetesimals are differentiated. This can be seen on Fig.~\ref{fig:cumdist_sims}, where the third quartile (84\%) occurs at Ca/Fe$=0.1$ for the model population and Ca/Fe$=0.24$ for the observed population. The model and the observed populations are consistent for low Ca/Fe ratios, but clearly diverge for Ca/Fe ratios above the average. Fig.~\ref{fig:hist_cf} uses the observed Ca/Fe ratios to predict the core mass fractions of the planetesimals accreted by the white dwarfs, using the inverse of Eq.~\ref{eq:cf}. This figure clearly shows that the Ca/Fe ratios observed would represent a significant overabundance of mantle-rich fragments compared to the simulated population. Here, we discuss other potential reasons for this divergence.


The observed population is by no means selected in an unbiased manner. The white dwarfs used here have been collated from the literature, where in general the objects with the best determined abundances are published. Even the large number of white dwarfs in \cite{Hollands2017} have not been selected in a uniform manner. Rather, they were identified in {\it SDSS DR 12} due to the presence of sufficient metal lines in the spectra to alter their broad-band magnitudes, such that they have redder u-g colours than the main-sequence in a u-g vs. g-r colour-colour diagram, rather than bluer like most white dwarfs. In addition to which, only white dwarfs where all three major elements Ca, Fe and Mg were detected are included. 

We can, however, consider the impact of requiring that both Ca and Fe are detected might have on the model population. Calcium is easier to detect than iron, yet, present in smaller quantities in planetary material. No calcium will be present in purely core fragments, which have, therefore been removed from the model population as undetectable. Fig.~\ref{fig:obs} shows that the fragments with low Fe abundances have high Ca/Fe ratios and thus, the non-detection of these fragments leads the observational sample to be biased towards low Ca/Fe ratios. This is the opposite trend to that seen in the observations. We, therefore, consider that we do not find strong evidence that observational biases are responsible for the overabundance of high Ca/Fe ratios in the observed population.

One tendency of the simulations, as discussed in \S\ref{sec:sims}, is to produce more mantle-rich fragments than core-rich fragments. This occurs as many gentle collisions chip off small pieces of mantle. Unfortunately the simulations are unable to resolve these small fragments and therefore, they are assumed to reaccrete onto all fragments uniformly. This leads to many fragments with core fractions slightly below average. However, if these fragments could be followed in detail, or if collisions were more violent, such as may occur in a system stirred by giant planets, it is plausible that  the distribution of core mass fractions may be skewed to contain more very mantle-rich fragments. Such fragments would produce a model population that tends to have more mantle-rich fragments and be more similar to that predicted from the observations, however, there will still be many more moderately mantle-rich fragments than extremely mantle-rich fragments. Fig.~\ref{fig:hist_cf} indicates how extreme this overabundance of high Ca/Fe ratios in the observed population is compared to the simulations, which makes it hard to explain with this model. The observations were sensitive enough to detect white dwarfs with low Ca/Fe ratios, as demonstrated by a handful of extreme examples, but these were found to be rare in \cite{Hollands2017}.

The simulations assume a single initial Ca/Fe ratio of Ca/Fe$=0.07$ ($C_f=0.35$) for all planetesimals. A range of initial Ca/Fe may be more appropriate across diverse exoplanetary systems, and indeed a range of Ca/Fe ratios are seen across chondritic meteorites in our Solar System (0.04-0.11 \cite{Wasson1988}, although some of the spread may be due to the effects of heating). We do not deem a range in the initial Ca/Fe as a likely explanation for the observed overabundance, as a large spread towards high Ca/Fe would be required.

The collision model did not include crustal differentiation. This has the potential to increase the number of fragments with high calcium abundances \citep{Carter2018}. Crustal stripping is particularly efficient at producing small fragments with high Ca/Fe ratios. Such fragments would also have altered abundances of other species. In addition to this, crustal material is generally a significantly smaller proportion of a planetary body's overall mass budget (\eg <0.5\% for Earth) and even when it is taken into account that smaller planetesimals may have deeper crusts, like Vesta, it is hard to envisage that crustal material accounts for such a large fraction of the population.

Processes related to heating at temperatures higher than 1,000K can, on the other hand, increase the Ca/Fe ratio of planetary bodies. Ca is more refractory than iron, with a 50\% condensation temperature of 1,660K compared to 1,357K \citep{Lodders2003}. This means that at typical temperatures occurring in the inner regions of proto-planetary discs, grains that condensed at temperatures between around 1,000K and 2,000K could contain more calcium-rich minerals than iron-rich minerals. Alternatively heating during the star's evolution on the giant branch could heat material to similar temperatures, leading to the removal of more iron-rich minerals. Such processes can only enhance, and not deplete, the Ca/Fe ratio of solids. We, therefore, propose that heating could lead to the overabundance of planetesimals with high Ca/Fe ratios accreted by white dwarfs, in a similar manner to \cite{Harrison2018} and assess the validity of this model in the next section.

\subsection{High temperatures lead to an increase in Ca/Fe}
\label{sec:hight}
In this section, we investigate whether a model in which some planetesimals accreted by white dwarfs have higher Ca/Fe ratios due to the effects of heat processing can explain the white dwarf observations.  The increased Ca/Fe ratio could have occurred due to incomplete condensation of iron-rich minerals during planet formation, or evaporation during formation or subsequent evolution, for example on the giant branch. The exact distribution of Ca/Fe is, thus, unknown. There is an additional complication that heating and differentiation can occur in the same planetesimals. We, therefore, decide not to focus on producing a detailed model of heating, but make a broad, all encompassing model that can tell us whether this explanation has the potential to be consistent with the population.  We, therefore, parametrise the distribution of Ca/Fe ratios due to heating as a normal distribution centred at $d_{\rm Ca/Fe}$, of width $\sigma_W$, where both $d_{\rm Ca/Fe}$ and $\sigma_W$ are model parameters. Clearly, such a model is sufficiently flexible to fit the observations, given appropriate values of $d_{\rm Ca/Fe}$ and $\sigma_W$, however, it can provide an indication of the fraction of the population for which heating is important.

We create a model in which a fraction, $f_{\rm hot}$, of the sample have undergone heating which leads to a cumulative distribution of Ca/Fe abundances, $k^{\rm hot}({\bf X_{\rm Ca/Fe}})$, given by the cumulative distribution function of a Gaussian centred on $d_{\rm Ca/Fe}$ and of width $\sigma_w$.  At the same time we consider that a fraction, $f_{\rm diff}$, of the sample are fragments of differentiated planetesimals, with a distribution of abundances that follow those of the simulations, $k^{\rm sims} ({\bf X_{\rm Ca/Fe}})$. Those white dwarf pollutants that are not differentiated, nor have experienced heating ($1-f_{\rm diff}-f_{\rm hot}$), have abundances that originate from the distribution of potential initial abundances, \ie from nearby stars, $ k^{\rm stars}({\bf X_{\rm Ca/Fe}})$. This leads to a model with four free parameters, $f_{\rm diff}$, $f_{\rm hot}$, $d_{\rm Ca/Fe}$ and $\sigma_w$. 
 We use this to calculate the model population, where $ k^{\rm model} ({\bf X_{\rm Ca/Fe}})$ is the cumulative distribution of Ca/Fe ratios in the model population, where: 
\begin{eqnarray}
 &k^{\rm model} ({\bf X_{\rm Ca/Fe}}) = f_{\rm diff}\, k^{\rm sims} ({\bf X_{\rm Ca/Fe}}) +\\\nonumber
&  (1-f_{\rm diff} - f_{\rm hot})\, k^{\rm stars}( {\bf X_{\rm Ca/Fe}}) + f_{\rm hot}\, k^{\rm hot} ({\bf X_{\rm Ca/Fe}}).
\label{eq:hot}
\end{eqnarray}
 We find the best-fit values of $f_{\rm diff}$, $f_{\rm hot}$, $d_{\rm Ca/Fe}$ and $\sigma_w$ by maximise the posterior distribution (Eq.~\ref{eq:like}) using the MCMC fitting routine of \citep{emcee} and assuming uniform priors of $0<f_{\rm diff}<1$, $0<f_{\rm hot}<1$, $0.07<d_{\rm{Ca/Fe}}<0.4$ and $0<\sigma_W<0.3$. The mid-point of the Gaussian is fixed to occur at high Ca/Fe ratios, \ie above Ca/Fe$=0.07$, equivalent to $C_f=0.35$, otherwise the priors are designed to be non-informative and encompass the full range of potential values for the parameters.

\begin{figure*}
\includegraphics[width=0.6\textwidth]{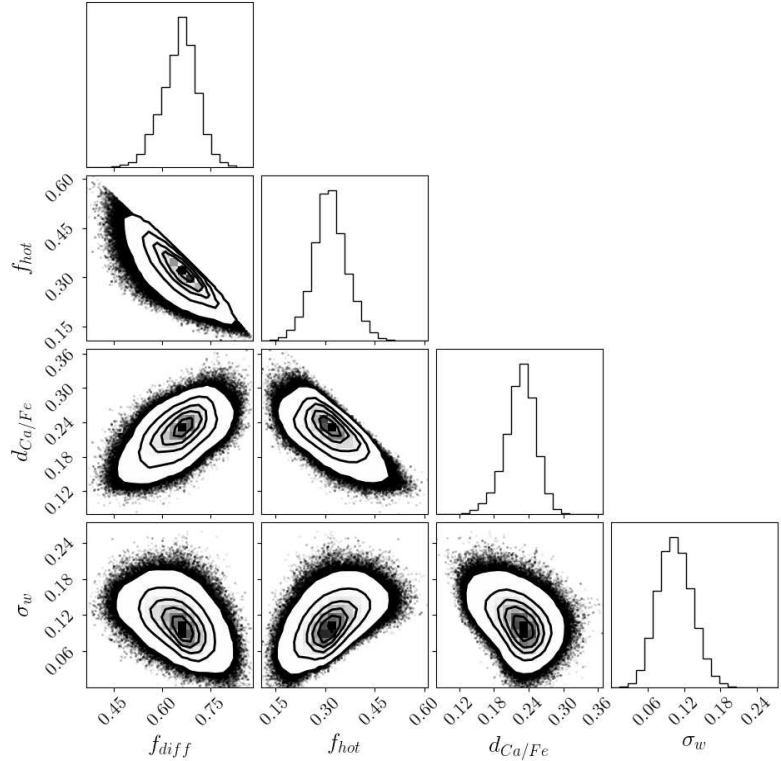}

\caption{The posterior probability distributions of each parameter in the empirical fit to observations, calculated by maximising the likelihood (Eq.~\ref{eq:hot}), assuming that a fraction $f_{\rm diff}$ of planetesimals are differentiated, whilst $f_{\rm hot}$ are subject to heating. Results are included for 4 high resolution {\it Calm} simulations. Plotted are individual walkers, with density contours overplotted at $(0.1,0.5,1.0,1.5,2.0) \sigma$, created using \citet{corner}.}
\label{fig:corner}
\end{figure*}

The results of the fitting procedure are listed in Table~\ref{tab:model}, whilst the posterior probability distributions of the four parameters are shown in Fig.~\ref{fig:corner}. Fig.~\ref{fig:corner} plots all 50 walkers at steps 1,000 to 2,000, in the 4 {\it Calm} simulations, with contours overlaid at $(0.1,0.5,1.0,1.5,2.0) \sigma$. The distributions show the existence of a clear best-fit solution.

 Fig.~\ref{fig:corner} shows there is a correlation between $f_{\rm diff}$ and $f_{\rm hot}$. Those models in which more exoplanetesimals are differentiated ($f_{\rm diff}$ is higher) show less effects of heating ($f_{\rm hot}$ is lower). This naturally makes sense as both parameters help to explain the excess of polluted white dwarfs with high Ca/Fe ratios. However, solutions for $f_{\rm diff}$ and $f_{\rm hot}$ all lie within defined ranges, with best-fit values of $66^{+4}_{-6}\%$ for $f_{\rm diff}$ and $f_{\rm hot}=31^{+5}_{-5}\%$ , averaged across all {\it Calm} simulations. The fraction of planetesimals that are differentiated is in-line with the model that focussed only on low Ca/Fe ratios (\S\ref{sec:fdiff}), the difference here being that a model in which all exoplanetesimals are differentiated is no longer such a good model for the data, which instead favours a significant fraction of exoplanetesimals to be influenced by the effects of heating.

For the heating model best-fit values of $d_{\rm Ca/Fe}=0.23^{+0.03}_{-0.03}$, $\sigma_{\rm w}=0.1^{+0.03}_{-0.03}\%$, averaged across all {\it Calm} simulations. In terms of the mid-point ($d_{\rm Ca/Fe}$) and width ($\sigma_w$) of the Gaussian, the majority of solutions indicate the addition of planetesimals with Ca/Fe ratios between 0.1-0.3 due to heating. These are very plausible values for objects where some iron rich minerals have been removed and the Ca/Fe ratio was fixed prior to differentiation. Such high Ca/Fe ratios are less likely to occur following differentiation, as they would require the removal of more iron than generally found in the body's mantle.



\begin{figure}
\includegraphics[width=0.48\textwidth]{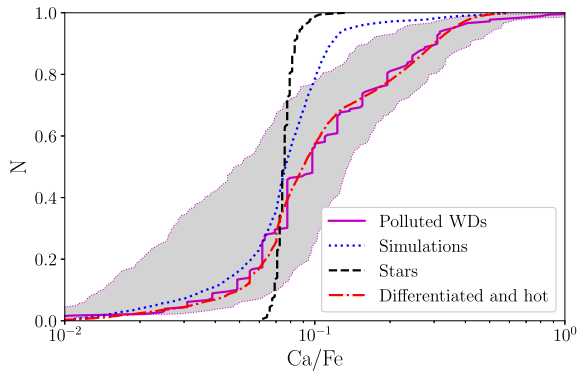}

\caption{The same as Fig.~\ref{fig:cumdist} now including the cumulative distribution of Ca/Fe abundances predicted from the model with both heating and differentiation (red dot-dashed line), compared to the blue dotted line which shows the {\it Calm 1} simulation. $f_{\rm hot}=31\%$, $f_{\rm diff}=66\%$, $d_{\rm Ca/Fe}=0.24$ and $\sigma_W=0.09$. }
\label{fig:cumdist_hot}
\end{figure}



\begin{table*}

\begin{tabular}{c c c c c c c}
\hline
\hline
Model  & $\chi^2$ & $\ln  \, \mathcal{L} $  & $f_{\rm diff}$ & $f_{\rm hot}$& $d_{\rm Ca/Fe}$& $\sigma_{\rm w}$\\
\hline
\hline

Stars & 3.5 & -180 \\


\hline

Diff: GT &  $          0.83 $ &  $     230     $ &  $ 98.4^{+1}_{-2}\% $ \\

Diff: Calm1 &  $         0.46 $ &  $  300        $ &  $ 98.2^{+1}_{-2}\%$ \\
Diff: Calm2  &  $         0.26 $ &  $ 330 $ &  $ 98.5^{+1}_{-3}\%$ \\

Diff: Calm3 &  $         0.43 $ &  $        300 $ &  $ 98.1^{+1}_{-3}\%$ \\
Diff: Calm4 &  $         0.42 $ &  $         305 $ &  $ 98.2^{1}_{-3}\%$ \\







\hline

Hot: GT  &  $         0.57 $&  $         330$ &  $         69^{+1}_{-1}\%  $ &  $         30^{+1}_{-1}\% $ &  $0.25\pm 0.01$ &  $         0.07\pm 0.02$\\

Hot: Calm1  &  $         0.13 $ &  $      370  $&  $         62^{+5}_{-6}\%  $ &  $         34^{+5}_{-5}\% $ &        $0.23\pm 0.02$       &$ 0.11\pm0.03$\\

Hot: Calm2  &  $         0.08 $&  $        374$ &  $         66^{+8}_{-7}\% $&  $        29^{+6}_{-8}\% $ &  $0.21\pm 0.03$ &  $  0.11\pm0.03$ \\


Hot: Calm3   &  $         0.16 $&  $         367$ &  $         67^{+3}_{-3}\%$ &  $         31^{+3}_{-3}\%$ &    $0.23\pm0.02 $ &           $0.09\pm 0.02$ \\

Hot: Calm4  &  $         0.09 $ &  $        373$&  $         66^{+5}_{-5}\%  $ &  $         31^{+4}_{-5}\% $ &      $0.23\pm 0.02$ &  $    0.11\pm0.03$ \\

\hline\hline

\end{tabular}

\caption{A table to show the results of comparing the model populations to the white dwarf observations. Best-fit parameters are shown, alongside reduced chi-squared (Eq.~\ref{eq:chisquared}) and likelihoods (Eq.~\ref{eq:like}).}
\label{tab:model}

\end{table*}

\section{Discussion}

\label{sec:discussion}

In this work we present a model in which the population of calcium and iron abundances observed in a sample of 179 polluted white dwarfs can be explained by the differentiation and collisional processing of a substantial fraction of the accreted planetesimals and the effects of processing at temperatures higher than 1,350K. We aim to constrain how frequently planetesimals accreted by white dwarfs are the collision fragments of differentiated bodies. In this section we discuss how robustly we can come to a conclusion.

Our null hypothesis was that the range of abundances observed in the planetary bodies accreted by white dwarfs resulted from a range of initial abundances present in the material out of which the planetary bodies formed. Using nearby stars as a proxy for this range of compositions, we show that this model is unlikely to explain the data. We deem it likely that differentiation, rather than other unknown processes, are responsible for the abundances in at least some white dwarfs due to the correlations observed in multiple siderophile species (core) or multiple lithophile species (mantle) which point towards segregation due to melting and differentiation \citep{Harrison2018, juraYoung2014}. In fact, those polluted white dwarfs in this sample with the lowest Ca/Fe ratios also have the lowest Mg/Fe ratios, pointing towards a common origin to the depletion of both lithophiles \citep{Hollands2018}. In a similar manner correlations between the abundances of multiple species linked by similar condensation temperatures, provides evidence for heat processing in individual objects \citep{Harrison2018}.

A model in which all white dwarf pollutants are fragments of differentiated bodies is consistent with the data, given the uncertainties. However, an excess of high Ca/Fe ratios in the observed sample compared to the model remains. We hypothesise that this excess can be explained by planetesimals that have suffered the effects of heat processing at temperatures between 1,000K to 2,000K. At such temperatures calcium-rich and iron-rich minerals exhibit different behaviours, with iron-rich minerals tending to be removed preferentially from the solid phase compared to calcium-rich minerals, which can lead to high Ca/Fe ratios in planetary bodies. We know from the Solar System that the effects of the depletion of moderately volatile elements, including trends featuring Ca and Fe are seen in meteoritic samples \citep[\eg][]{Palme2003treatise}. It, therefore, seems probable that such processes would have occurred in the planetesimals accreted by white dwarfs, long prior to their accretion onto the white dwarfs.

The key question regards the fraction of exoplanetesimals that are differentiated. Based on the model analysis, we consider that between 60\% and 100\% of exoplanetesimals are differentiated, with a most likely value of $66^{+4}_{-6}\%$. The best way to improve upon these conclusions would be to increase the sample size, particularly the sample size of white dwarfs with precisely determined abundances of at least Ca and Fe. For example, there are less than 10 white dwarfs with low Ca/Fe$<0.07$ and high $S/N$ ($S/N>10$) from which the most can be learnt about the fraction of planetesimals that are differentiated. This is not yet large enough to smooth out the effects of small number statistics.

\subsection{Limitations of the model}
\label{sec:limitations}

The current model suffers from many limitations, which will be discussed here. However, we note that the limitations of the model must be considered in the context the size of the data sample with sufficiently precise abundance determinations and our lack of knowledge regarding any biases in its selection. The most significant weakness of this work is that we focus solely on Ca and Fe. Whilst this allows us to process a larger data sample in the same manner, we are not taking advantage of the full information available for each white dwarf. Ca and Fe are a very useful pair of elements as the effects of differential sinking in the white dwarf atmosphere can be ignored. Conclusions regarding individual objects will be found in Harrison et al, in prep.

The model is relatively simplistic. We create a model of collision fragments based on the collisional history of a single planetary system. Whilst the white dwarf pollutants necessarily originate from a wide range of systems, it is not feasible to simulate even a small range of the potential architectures. We do, however, predict that the range of core mass fractions produced will vary in width and height, rather than significantly in form. We hypothesise that the generic form, whereby collisional evolution produces produces a centrally peaked distribution of Ca/Fe ratios, or indeed core mass fractions, with fewer fragments possessing extreme values, will be consistent across a wide range of system architectures. 

The main conclusions of this work are based on the existence of many more bodies with Ca/Fe (core mass fractions) close to a central (original) value than extreme values. In fact, the fraction of exoplanetesimals that must be differentiated would only increase, if the distribution were more centrally peaked, a plausible consequence of more dramatic collisional evolution, as seen in the GT simulation (see Fig.~\ref{fig:hist_sims}). If the distribution of Ca/Fe ratios were spread more broadly in systems with particular collision histories, it is plausible that the fraction of exoplanetesimals that are differentiated has been overestimated. However, the number of white dwarf pollutants with extreme Ca/Fe ratios limits any potential reduction.

We sample the Ca/Fe ratios produced in a single planetary system at a single epoch. We can consider this to be a reasonable approximation to the collisional evolution at larger orbital radii on longer timescales, as discussed in \S\ref{sec:diff}. However, necessarily we cannot sample a full range of systems at a full range of applicable epochs. We can, however, justify the sampling of a single system at a single epoch, as the collisional evolution tends after a period of initial evolution to a relatively steady distribution of Ca/Fe in the bodies considered. This means that, if the white dwarf pollutants sample systems at a range of orbital radii, we anticipate that the distribution in Ca/Fe may not depend strongly on orbital radii, for those systems where collisional equilibrium is established. This does, however, require further detailed investigation.

We assume that each white dwarf has accreted a single fragment. If instead, multiple fragments were important, this could potentially act to smooth the distribution of Ca/Fe ratios \citep{Turner2019} and we would expect to see fewer examples of white dwarf pollutants with extreme abundances. This would make it harder to conclude that any white dwarf pollutants are not the fragments of differentiated exoplanetesimals. It is, however, potentially possible that white dwarfs accrete differentiated bodies in a biased manner, for example accreting first material from the mantle, followed by material from the core.

We ignore crustal fragments, which do have the potential to alter the Ca/Fe ratio. This would likely mean that some pollutants with high Ca/Fe may result from crustal fragments, rather than heating, thus reducing $f_{\rm hot}$, whilst leaving $f_{\rm diff}$ unaffected. This is an important avenue for future investigations.


The heating model is very simplistic and presented merely to show that heating is a plausible explanation for the overabundance of white dwarfs with high Ca/Fe ratios, rather than as a precise distribution of the Ca/Fe ratios likely to exist in planetesimals following the effects of temperatures higher than 1,000K. We refer the interested reader to \citep{Harrison2018} for a more detailed description of the Ca/Fe ratios that may result from the effects of heating on exoplanetesimals.





\subsection{The implications of the results for exoplanetary systems }

We have shown that many white dwarf pollutants are likely to be the collision fragments of planetesimals that differentiated to form a core and a mantle. The implications of this conclusion depends on the size of planetesimals that are accreting onto white dwarfs, which remains an open question. Whilst we can measure the mass of material currently in the atmosphere of the white dwarf, which in some cases is greater than the mass of Ceres \citep{Raddi2015}, exactly how this relates to the size of the body accreted, or the size of the parent body, if the pollutant is a collision fragment, remain unclear.

If the white dwarf pollutants are the collision fragments of parent bodies larger than around $>1,000$\,km in diameter, their differentiation can be explained by the heating imparted by impacts occurring during planet formation \citep{Davison2010}. This would imply that a large proportion of exoplanetary systems include collisionally evolved populations of Pluto-sized bodies, something which does not seem to be the case within our own Solar System, for example where $D>100$\,km bodies in the asteroid belt are mostly primordial \citep[\eg][]{Bottke2015}. However, white dwarf planetary systems have the potential for dynamical instabilities to have occurred due to mass loss post-main sequence \citep[\eg][]{Veras_twoplanet_2013, Mustill_threeplanet}.

On the other hand, the planetesimals accreted by white dwarfs could in general be collisional fragments of bodies smaller than $1,000$\,km. In this case it is harder to understand what powers their differentiation unless there is a significant source of heating present from short-lived radioactive nuclides in most exoplanetesimals, as suggested by \citep[\eg][]{Jura2013, Young2014}. Such a theory would support a model in which the presence of these nuclides leads to triggered star formation, such that all planetary systems initially have a large budget of such nuclides \citep[\eg][]{Boss2008, Li2014}. In addition to which it would imply that most white dwarf pollutants originate from sufficiently close to the star that planetesimal growth had made it to large enough sizes to differentiate prior to the decay of any short-lived radioactive nuclides present in the planet forming material \eg $^{26}$Al with its half life of 0.7Myr. 


The requirement for the collisional evolution of large planetesimals, whether they are of the order of $\sim 10$\,km in size, or much larger, in most white dwarf planetary systems implies that most of these systems possess debris belts significantly more massive than our own Solar System's asteroid or Kuiper belt. This agrees with observations of main-sequence debris discs, which find that, for example, $\sim30\%$ of main-sequence A stars have a debris belt detectable with {\it Spitzer} \citep{su06, wyattreview}, whilst our asteroid and Kuiper belt lie orders of magnitude below the detection limit. Another possibility is that the dynamical instabilities that lead to white dwarf pollution \citep[\eg][]{bonsor11, DebesSigurdsson, debesasteroidbelt} lead to collisional processing in large bodies. The most likely scenario is that in some systems, the white dwarf pollutants result from the collisional processing of very large bodies (>1,000\,km) which are differentiated, whilst in other systems, the presence of short-lived radioactive nuclides leads to the differentiation of smaller bodies.

We have also shown that heat processing above 1,350K of some material in about a third of the white dwarf pollutants can explain their Ca and Fe abundances. Such temperatures are only reached during planet formation interior to 1.5\,au in a \cite{Chambers2009} proto-solar nebula \citep{Harrison2018}. On the other hand heating during the giant branch evolution suggests that a $3M_\odot$ star can reach a luminosity of $16,000L_\odot$ at the tip of the AGB \citep{sse}, which would imply an equilibrium temperature of $>1,350$K for bodies interior to 6\,au. If this explanation for the Ca/Fe abundances is correct, this implies that about a third of white dwarf pollutants originate from planetesimal belts within a few au of their host stars. There is clear evidence that some white dwarf pollutants which have retained volatiles such as water originated from planetesimal belts outside of the ice line \citep{Farihi2011_water, Raddi2015, Malamud2016}. Thus, in combination with the need for heating, white dwarf abundances imply that a large spread in the original radii of white dwarf pollutants must exist.


\section{Conclusions}
\label{sec:conclusions}
Abundances of calcium and iron in the atmospheres of white dwarfs can be used to study the differentiation of exoplanetesimals. We use a sample of 179 white dwarfs collated from the literature, where both Ca and Fe are detected, to show that the distribution of Ca/Fe ratios are unlikely to occur as a result of a distribution in the initial compositions (Ca/Fe) at the start of planet formation. We hypothesise that, instead, a fraction of the white dwarfs have accreted the fragments of differentiated exoplanetesimals.

If exoplanetesimals differentiate, collisions can lead to fragments with a range of core mass fractions (Ca/Fe). We present results from a set of N-body simulations in which the fate of core and mantle material during collisions is traced separately. These simulations show that the distribution of core mass fractions in the remnant planetesimals is always dominated by values close to the core mass fraction of the parent bodies, with a few extremely core-rich or mantle-rich fragments. This means that whilst it is easy to conclude that white dwarf pollutants with extreme Ca/Fe ratios are likely to be core-rich or mantle-rich fragments, for every extreme observation, we anticipate many more white dwarf pollutants with less extreme Ca/Fe ratios.

Using the simulation results, we created a model population of planetesimals that could accrete onto polluted white dwarfs. We use this model to show that the observed range of Ca/Fe ratios in polluted white dwarfs is consistent with all polluted white dwarfs having accreted the collision fragment of a differentiated planetesimal, however, there is an overabundance of polluted white dwarfs observed with high Ca/Fe ratios. We suggest that this is unlikely to be an observational bias and more likely a result of processing at temperatures between 1,000K and 2,000K during the formation or subsequent evolution of the accreted planetesimals. In this case, our best-fit model finds that $31^{+5}_{-5}\%$ of planetesimals accreted by white dwarfs have increased Ca/Fe due to the effects of heating, whilst $66^{+4}_{-6}\%$ are the fragments of differentiated planetesimals.

These results imply that the collisional evolution of
large planetesimals (at least larger than tens of kilometres)
is a typical feature of exoplanetary systems, in line with
observations of debris discs around main-sequence stars. The population of white dwarf pollutants suggest that differentiation occurs commonly in exoplanetesimals and that either short-lived radioactive nuclides are present in many exoplanetary systems, or white dwarf pollutants are typically the collision remnants of planetary bodies sufficiently large that impacts during planet formation can lead to differentiation. We highlight the need for a larger sample of white dwarfs with precisely determined abundances to investigate further the geological process of differentiation in exoplanetary systems.

\section*{Acknowledgements}
AB acknowledges funding from a Royal Society Dorothy Hodgkin Fellowship. JHDH acknowledges an STFC studentship. BTG was supported by the UK STFC grant ST/P000495. PJC acknowledges support from UC Office of the President grant LFR-17-449059. The research leading to these results has  received funding from the European Research Council under the European Union's Horizon 2020 research and innovation programme no. 677706 (WD3D)

\section*{Appendix}

\secondtable

\bibliographystyle{mn}

\bibliography{ref}


\bsp	
\label{lastpage}
\end{document}